\documentclass[12pt]{iopart}
\usepackage{times}
\usepackage{mathptmx}
\usepackage{graphicx}

\begin{document}
\title{Probing two-subband systems in a quantizing magnetic field  
with non-equilibrium phonons} 
\author{M.E. Portnoi, K.V. Kavokin}
\address{School of Physics, University of Exeter, Exeter EX4 4QL, UK}
\author{V.M. Apalkov}
\address{Department of Physics, University of Utah, Salt Lake City, UT 84112,
USA}
\begin{abstract}
We propose to use phonon absorption spectroscopy to study many-body
gaps and phases of two-subband heterostructures in the quantum Hall 
regime. Implications of the spin-orbit interaction for phonon 
absorption in this system are considered.  
\end{abstract}

Phonon spectroscopy is a powerful experimental technique that has
been successfully used in the study of the properties of a two-dimensional
(2D) electron gas in the quantum Hall regime [1]. It allows probing of the
states with non-zero momentum, which are not accessible by either cyclotron
resonance or photoluminescence methods. We propose to use acoustic-phonon
spectroscopy to study many-body effects in a two-subband quasi-2D system,
formed at a single semiconductor heterojunction and subjected to a
quantizing magnetic field. We show that the formation of many-body gaps and the
existence of different spin phases in this system result in a striking
difference in the magnetic field dependence of the phonon absorption rate at
different filling factors.

Acoustic-phonon scattering of 2D electrons in a quantizing magnetic field is
strongly suppressed due to the dispersionless structure of the Landau levels
(LLs). In a two-subband system, however, a sharp enhancement of the
electron-phonon interaction occurs when the two LLs corresponding to
different size-quantization subbands (we consider the second LL of the first
subband and the first LL of the second subband only) approach one another,
and the separation between these energy levels, $\Delta =\Delta _{12}-\hbar
\omega _{c}$, becomes close to $\hbar s/l$. Here $\hbar \omega _{c}$ is the
cyclotron energy, $\Delta _{12}$ is the intersubband splitting, $s$ is the
velocity of sound and $l$ is the magnetic length. This electron-phonon
interaction enhancement should manifest itself in the magnetic field
dependence of the phonon absorption rate and dissipative conductivity. At
filling factor slightly greater than two, when the electron-electron
interaction between the particles belonging to either of the intersecting
LLs can be neglected, a double-peak structure in the magnetic field
dependence of the phonon-mediated transport properties of the two-subband
system has been predicted [2]. With increasing electron density, the repulsion
between the levels due to electron-electron exchange interactions opens a
gap in the interlevel excitation spectra. At filling factor $\nu =3$ this
gap exists at all values of the bare level splitting $\Delta $ and can well
exceed $\hbar s/l$. As a result, the double-peak structure transforms into a
single-peak one (see Figure 1). 
By far the most interesting case is when the electron
filling factor $\nu =4$. When the cyclotron energy is close to the
intersubband splitting the system can be mapped onto a four-level electron
system with an effective filling factor $\nu ^{*}=2$. The ground state is
either a ferromagnetic state or a spin-singlet state, depending on the
values of the interlevel splitting and Zeeman energy $\Delta _{Z}$. The
electron-electron interaction renormalizes the phase boundaries, which
results in the existence of a ferromagnetic phase even for 
$\Delta _{Z}=0$. 
The excitation spectrum has a gap for any value of the interlevel
splitting $\Delta $ and Zeeman splitting $\Delta _{Z}$. This results in
strong suppression of the electron-phonon interaction. The rate of phonon
absorption by the considered quasi-2D electron system at $\nu =4$ again has 
a double-peak structure as a function of level splitting and a steplike
structure as a function of Zeeman splitting (see Figure 2).

In the absence of the spin-orbit interaction, the transitions between levels
with different spin indices are forbidden because the electron-phonon
interaction conserves the electron spin. The spin-orbit interaction allows
such transitions. Matrix elements of the spin-orbit interaction Hamiltonian
$H_{SO}$ between levels differing by subband number, LL number, and spin
projection on the normal to the structure plane, are given by following
expressions [3]:
\begin{eqnarray}
\left\langle \eta ,n,k_{x},-1/2\left| H_{SO}\right| \mu,
n-1,k_{x},+1/2\right\rangle  &=&iC\frac{1}{l_{H}}\sqrt{2n},  \nonumber\\
\left\langle \eta ,n,k_{x},+1/2\left| H_{SO}\right| \mu,n-1,k_{x},
-1/2\right\rangle  &=&A\frac{1}{l_{H}}\sqrt{2n},  \nonumber\\
\left\langle \eta ,n-1,k_{x},-1/2\left| H_{SO}\right| \mu,n,k_{x},
+1/2\right\rangle  &=&A\frac{1}{l_{H}}\sqrt{2n},  \nonumber\\
\left\langle \eta ,n-1,k_{x},+1/2\left| H_{SO}\right| \mu,n,k_{x},
-1/2\right\rangle  &=&-iC\frac{1}{l_{H}}\sqrt{2n},  \label{1}
\end{eqnarray}
where $H_{SO}=\left( \alpha \hbar ^{3}\left( 2m^{3}E_{g}\right)^{-1/2}
\mathbf{K} + \beta \left[ \mathbf{k\times }\nabla V\right] \right)
\cdot \mathbf{S}$, $~\mathbf{K} =\left
[k_{x}(k_{y}^{2}-k_{z}^{2}),k_{y}(k_{z}^{2}-k_{x}^{2}),k_{z}
(k_{x}^{2}-k_{y}^{2})\right]$, 
$\eta $ and $\mu $ numerate subbands, $n$ is the LL number, and $\alpha $ and 
$\beta $ are phenomenological spin-orbit constants of the semiconductor [4].
The expressions for $A$ and $C$ are given by 
\begin{eqnarray}
A=\alpha \hbar ^{3}\left( 2m^{3}E_{g}\right) ^{-1/2}\int\limits_{-\infty}
^{+\infty }\Psi _{\mu }(z)\frac{d^{2}\Psi _{\eta }(z)}{dz^{2}}dz, 
\nonumber\\
C=\beta
\int\limits_{-\infty }^{+\infty }\Psi _{\mu }(z)\Psi _{\eta }(z)\frac{dV_{QW}
}{dz}dz, \label{2}
\end{eqnarray}
where $V_{QW}$ is the quantum-well confinement potential. These matrix
elements cause an admixture of states with opposite spin direction to both
the initial and final states for the phonon absorption. As a result, the phonon
absorption rate at $\nu =4$ never becomes zero; in the region between peaks,
it is of the order of $(M/\Delta _{g})^{2}$ times its value at the peak of
absorption, where $M$ is the corresponding matrix element of the spin-orbit
interaction (Eq.(1)), and $\Delta _{g}$ is the many-body energy gap.

In the absence of many-body effects, the inter-subband spin-orbit coupling
results in anticrossing of levels with different spin, subband and LL
indices. The resulting energy gaps can be measured by high-resolution
spectroscopic techniques (optical or spin/cyclotron resonance spectroscopy),
as has been suggested for one-subband systems in tilted magnetic fields
[5].

This work was supported by the UK EPSRC.

\begin{center}
\begin{figure}[tbp]
\begin{center}
\includegraphics[width=13.5cm,keepaspectratio]{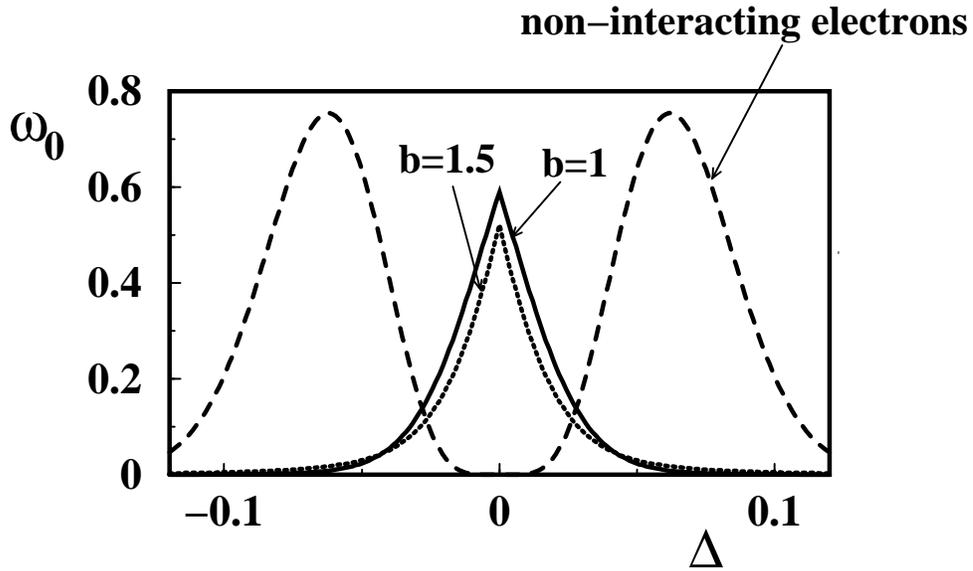}
\end{center}
\caption{The phonon absorption rate per phonon, $\omega_0=
\omega_{abs}/n(q_0) $, as a function of bare level separation, 
$\Delta=\Delta_{12} - \hbar \omega_c$, at filling factor $\nu=3$ (solid and
dotted lines) and for the case of non-interacting electrons (dashed line).
The results for $\nu=3$ are shown for two different values of the
Fang-Howard parameter (inverse quantum well width) $b=1$ and $b=1.5$ in
units of inverse magnetic length $l^{-1}$. The absorption rate $\omega_0$ is
in units of $10^{10}~\mathrm{s}^{-1}$; $~\Delta$ is in Coulomb units ($
\varepsilon_C=e^2/\kappa l$).}
\end{figure}
\begin{figure}[tbp]
\begin{center}
\includegraphics[width=13.5cm,keepaspectratio]{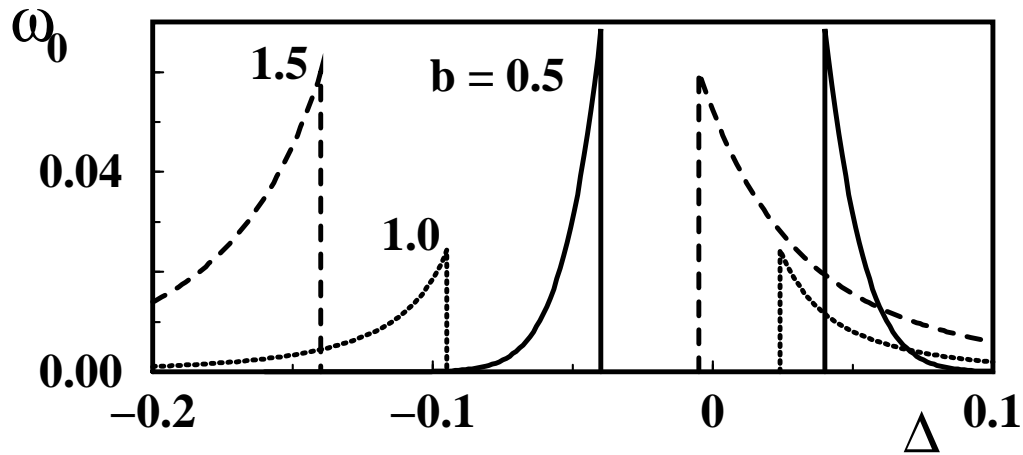}
\end{center}
\caption{The phonon absorption rate $\omega_0$ at filling factor $\nu=4$ as
a function of interlevel separation $\Delta$ for the case of zero Zeeman
splitting, $\Delta_Z=0$, for three values of parameter $b$. The data for $b=$
1.0 and 1.5 are multiplied by 10. The absorption rate $\omega_0$ is in units
of $10^{10}~\mathrm{s}^{-1}$; $\Delta$ is in units of $\varepsilon_C$.}
\end{figure}
\end{center}

\section*{References}
\begin{thebibliography}{9}
\bibitem{1}U. Zeitler, A.M. Devitt, J.E. Digby, C.J. Mellor, A.J. Kent, 
K.A. Benedict, and T. Cheng, Phys.\ Rev.\ Lett.\ \textbf{82},
5333 (1999), and references therein.
\bibitem{2} V.M. Apalkov and M.E. Portnoi, Physica E\ \textbf{12}, 470 (2002).
\bibitem{3} K.V. Kavokin and M.E. Portnoi, to be published.
\bibitem{4} G.L. Bir and G.E. Pikus, \textit{Symmetry and strain-induced 
effects in semiconductors} (Wiley, New York, 1974).
\bibitem{5} V.I. Falko, Phys. Rev. B \textbf{46}, 4320 (1992).
\end{thebibliography}
\end{document}